\documentclass[letterpaper]{article}
\usepackage{amsmath,amssymb,graphicx,enumerate,footnote}
\usepackage{verbatim,graphicx,fullpage,url}
\usepackage{graphicx,color,verbatim,enumerate}
\usepackage[firstpage]{draftwatermark}
\SetWatermarkText{PREPRINT}
\SetWatermarkLightness{0.9}
\SetWatermarkScale{4}
\begin{document}
\title{Is Stack Overflow Overflowing With Questions and Tags}
\author{Ranjitha R K and Sanjay Singh \thanks{Sanjay Singh is with the Department of Information and Communication Technology, Manipal Institute of Technology, Manipal University, Manipal-576104, INDIA, E-mail: sanjay.singh@manipal.edu}}
\maketitle
\begin{abstract}
Programming question and answer (Q\&A) websites, such as Quora, Stack Overflow, and Yahoo! Answer etc. helps us to understand the programming concepts easily and quickly in a way that has been tested and applied by many software developers. Stack Overflow is one of the most frequently used programming Q\&A website where the questions and answers posted are presently analyzed manually, which requires a huge amount of time and resource. To save the effort, we present a topic modeling based technique to analyze the words of the original texts to discover the themes that run through them. We also propose a method to automate the process of reviewing the quality of questions on Stack Overflow dataset in order to avoid ballooning the stack overflow with insignificant questions. The proposed method also recommends the appropriate tags for the new post, which averts the creation of unnecessary tags on Stack Overflow. 
\end{abstract}
%
%

\section{Introduction}
Software development is a complex activity that involves many concepts such as the usage of APIs, interface defined, bugs fixed, architectural designs etc. In order to work with these aspects, the programmers seek different sources of information to accomplish these tasks. To lift the burden of solving problems on their own, code examples help them in a much easier way. Thus the developers rely on different sources of knowledge to find answers to their problems. One of the known programming question and answer website to obtain solution to their problems is Stack Overflow (SO). 
\par
Stack overflow is an online platform for developers to post their programming questions, provide answers to the existing questions and find solution to their difficulties faced during programming. A developer must add tags while posting a question to help other users to find out what the question is about. If the answer provided by any user gives solution to the problem faced by the questioner, that answer can be selected by the questioner which is called the \emph{accepted answer} to that question. Different members of the site can vote on questions and answers. The positive votes called the \emph{upvote} and negative votes called the \emph{downvote}, which shows how helpful that question/answer was for the users.
\par
The score of a question/answer is determined by the difference between the number of up/down votes. Based on the different activities of each user on Stack Overflow such as posting questions or answers, voting on them, posting comments, etc. their reputation score increases which help them build their reputation on stack overflow website. Greater the reputation values, more the capabilities for a member on stack overflow like deletion of questions/answers, closing questions, etc.  
\par
Each post contains meta data such as title, body, creation date, post type id, view counts, answer count, comment count etc. Title contains the short detail about the questions being posted by the developer. Body contains the complete details about the question which may also include the code snippets. As of early August 2010, Stack Overflow has a total of 300k registered users who asked 833k questions and the site served 7.8 million monthly visitors \cite{Mamykina}. According to the stack overflow data dump of September 2013 provided by the Stack Exchange network, Stack Overflow stores more that 5.5M questions, 10.2M answers and a community counting more than 2.3M users \cite{Ponzanelli}.
\par
Over a period of time, these websites turn into knowledge repositories of software engineering. Thus by analyzing and understanding this knowledge repository we obtain key insights to the use of specific technologies by the developers and trend of developers discussions. This further helps us in better understanding the thoughts and needs of developers. The methodology used for analyzing such a knowledge repository is Latent Dirichlet Allocation (LDA) \cite{Blei2}, a statistical topic modeling technique which automatically discover the main topics present in developers discussion. 
\par 
Once the stack overflow data set is extracted, we consider only the body part of each post as it contains the most text contents. We need to preprocess the text contents by discarding the code snippets, removing all the \texttt{html} tags and stop words. LDA is applied to the preprocessed data. The result of LDA is the number of topics to which each post is related. We have also analyzed the trends of the topics and the interacting patterns between the topics. This is followed by identifying related topics to the new questions entered by the developers and suggesting tags based on those topics. In addition to this, we analyze the quality of questions in our stack overflow dataset using the score value of each post. We also analyze the quality of questions related to each topic discovered by LDA. The main aim of identifying the quality of questions is to maintain the quality and growth of stack overflow website.

\section{Related Work}
\textit{General Q\&A Websites}: Previous work has focused on analyzing general Q\&A websites based on user's social interactions. Gyongyi et al. \cite{GGPG} have analyzed several aspects of user behavior in Yahoo! Answers, a Q\&A website for the general public. The authors use the number of questions and answers in each predefined top level category to determine the popularity of each category. Adamic et al. \cite{Adamic} have also analyzed Yahoo! Answers to cluster the top-level categories into three broader categories using both content and user interactions. In contrast to these efforts, instead of using existing tags, we use a statistical topic model, LDA, to automatically discover topics from the textual content of the posts and employ temporal measures to identify a topic's popularity over time.
\par

\textit{Stack Overflow}: Treude et al.\cite{Treude} have analyzed Stack Overflow to categorize its questions and identify its design features \cite{Mamykina}. Treude et al. \cite{Treude} have analyzed Stack Overflow to find topics and to categorize the question into distinct types, such as 'how-to', 'discrepancy', etc. They apply their analysis to 15 days worth of posts, using user-created tags to identify the topics. They manually code the questions based on a random sample of the data. In contrast, we apply our automated method, based on LDA, to 9 month's worth of posts and use tags as a secondary basis to deduce trends of different technologies under broader topic categories. 
\par
Mamykina et al. \cite{Mamykina} identify the core design features that led to the popularity of Stack Overflow: the reputation system based on points, the strong involvement of the design team with the community, and the single-domain focus. Further, the authors categorize users based on their frequency of activity on Stack Overflow, for example, community activists, and low profile users. Instead of user activity, we focus on the textual content generated by the users in order to extract the major topics of discussion.

\textit{Other Social Platforms}: Works have been reported on the analysis of other datasets and social platforms, such as code search engine usage logs and developer blogs, to find out the topics in which developers are interested. In particular, Bajracharya and Lopes \cite{Bajracharya} analyze the log of a popular code search engine to discover major code search topics. Similar to our technique, they apply LDA on the usage log of the code search engine. Some of the topics found by their analysis are aligned with our findings, for example, data structures, files, GUI, networking, parsing/compiling, security, and string. 
\par
However, their analysis is based on a specific group of developers, namely Java programmers. In contrast, our analysis takes into account developers using a myriad of different programming languages and platforms. Further, we analyze both questions and answers, whereas the aforementioned work analyzes only the question content (i.e., search queries). Moreover, their study is focused on a specific need of developers: finding source code examples for a particular problem. In contrast, our analysis of a community based Q\&A website addresses developer needs from a broader perspective.
\par
In another study, Pagano and Maalej \cite{Pagano} analyze the blogging activity of developers using topic models to find topics in those blogs. In particular, they analyze blogs written by developers who are active committers to a certain code base (e.g., Eclipse, PostgreSQL, GNOME, and Python).

\section{Methodology}
\begin{figure}[bpht!]
		\begin{center}
				 \includegraphics[height=5.8cm]{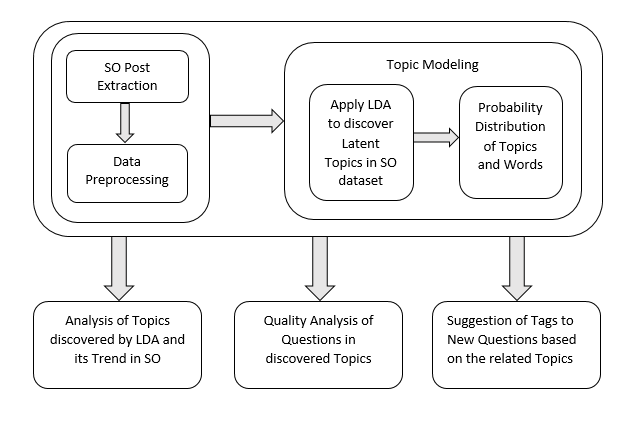}
				  \caption{An overview of proposed method}
				  \label{fig:f1}
		\end{center}
	\end{figure}

The tags are provided by the developers while posting the questions; currently Stack Overflow uses those tags to categorize each question posts. But the drawback of these tags are that most of the time it is erroneous and inconsistent which leads to Tag Explosion. To overcome this problem we use topic modeling technique LDA. Figure \ref{fig:f1} shows various steps of our proposed method. First, we extract the posts from Stack Overflow dataset. Secondly, we preprocess the extracted posts data. Third, we apply topic modeling technique LDA to the preprocessed data to overcome the problem of tag explosion. Finally, we analyze the output of LDA.

\subsection{Data Extraction}
We extract the posts \texttt{.xml} file from the Stack Overflow data dump, which contains all the user posts i.e., questions and answers from 31st July 2008 to 27th March 2009 on Stack Overflow. Each individual post is considered as a separate document. Thus total number of documents created are 513136 out of which there are 111871 question posts and 401265 answer posts. We consider both question and answer posts because most of the text contents are present in answer posts and we need to discover the relationship between question topics and answer topics. 
\par 
Each extracted post includes other details like creation date, post type, ID, user defined tags, etc. For answer posts there is a pointer to the question it is answering and for question posts there is a pointer to all its answers. 

\subsection{Data Preprocessing}
Once the Stack Overflow data has been extracted, we preprocess the extracted posts in four steps. First, we discard all the code snippets present in the posts. Since all code snippets contain similar programming language syntax and keywords, these do not help topic models to find useful topics. Second, we remove all the \texttt{html} tags for example $<b>$, $<a href="\ldots">$ etc. Third, we remove common English-language stop words such as "a", "the", "is" etc. Finally, we apply Porter stemming algorithm \cite{stemming}, which maps words to their base form, for example, "Implementation" and "Implementing" both get mapped to "Implement".

\subsection{Topic Modeling} 
Topic modeling is a suite of algorithms which help to discover and annotate large archives of documents with thematic information. Topic modeling algorithms are statistical methods that analyze the words of the original texts to discover the themes that run through them, how those themes are connected to each other, and how they change over time \cite{Blei1}. In this paper, we use one of the popular topic modeling technique called Latent Dirichlet Allocation (LDA) \cite{Lecture1}\cite{Lecture2}. 
\par
\begin{figure}[bpht!]
		\begin{center}
				 \includegraphics[height=3.5cm]{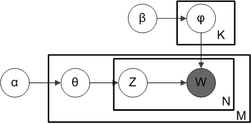}
				  \caption{Plate Notation of LDA \cite{plate}}
				  \label{fig:f2}
		\end{center}
	\end{figure}

 
LDA is a generative model for randomly generating observable data, given some hidden parameters \cite{Blei2}. LDA is a model for collection of discrete data such as text corpora. LDA provides probability distribution of topics over words in the corpus and probability distribution of documents over the discovered topics. In LDA, each document may be related to a mixture of various topics. LDA creates topics when it finds set of words that tend to co-occur frequently in the documents of the corpus. The plate notation of LDA is shown in Fig.\ref{fig:f2}.

In plate notation, the boxes are plates representing replicates. The outer plate represents documents which is denoted by $M$, while the inner plate represents the number of topics and words denoted by $K$ and $N$ respectively, which are distributed over documents \cite{Blei2}. Each document may be related to one or more topics and each word in the discovered topics will have its own probability. The calculated probability of each word $w$ over topics $z$ and probability of each topic over documents are represented by $\varphi$ and $\theta$ respectively. The $\alpha$ and $\beta$ are the parameters of Dirichlet distribution on per-document topic distribution and per-topic word distribution respectively. Here, their value is set to 0.01 and it can be set to any value between 0 and 1.

\subsubsection{LDA Implementation}
LDA uses Gibbs Sampling Algorithm \cite{Lecture} to infer the topics from the Stack Overflow data. Here we make use of The Stanford Topic Modeling Toolbox (TMT) \cite{tmt} for the implementation of LDA.

\subsubsection{Number of Topics}
The number of topics is denoted by $K$. It is a user specified parameter that controls the granularity of the discovered topics. If the value of $K$ is large, it produces more detailed topics and if the value is small, it produces more general topics. Here we aim for topics of medium granularity, so we set $K$ value to 10.

\section{Metrics and Analysis}
LDA discovers $K$ topics, $z_{1},\ldots,z_{k}$. The distribution of a particular topic $z_{k}$ in document $d_{i}$ is denoted as $\theta(d_{i}, z_{k})$. Note that $\forall i, k: 0 \leq \theta(d_{i}, z_{k})\leq 1 $ and $\forall i: \sum_{1}^{k} \theta(d_{i}, z_{k}) = 1$. Then, we define a threshold, $\delta$, to indicate whether a particular topics is "in" a document. A document $d_{i}$ can have between 1 to 5 dominant topics each with memberships of 0. Thus, by using the $\delta$ threshold as a membership cutoff, we keep only the main topics in each document and discard the probabilistic errors.

\subsection{Topic Share}
We define the overall \textit{share} \cite{anton} of a topic $z_{k}$ across all posts as,
\begin{equation}
share(z_k) = \frac{1}{|D|} \sum_{\substack{d_i\in D\\ \theta(d_i, z_k)\geq \delta}} \theta(d_i,z_k)       
\end{equation}
where $D$ is the set of all posts in our dataset. The share metric measures the proportion of posts that contain the topic $z_{k}$. For example, if a topic has a share metric of 10\%, then 10\% of all posts contain this topic. The share metric allows us to identify the major discussion topics in our Stack Overflow dataset.

\subsection{Topic Relationships}
We have to determine the relationship between topics in questions and topics in corresponding answers. We consider a discussion as a single question post along with its answer posts. We define the relationship \textit{rel} \cite{anton} between two topics $z_{q}$ and $z_{a}$ in one discussion as,
\begin{equation}
rel(z_q,z_a) = \sum_{\substack{d_i\in Q, d_j\in A(d_i)\\ \theta(d_i,z_q)\geq \delta\\ \theta(d_j,z_a)\geq \delta}} \theta(d_i,z_q)\times\theta(d_j,z_a)    
\end{equation}
where $Q$ is the set of all question posts and $A(d_{i})$ is the set of all answer posts related to question $d_{i}$.

\subsection{Topic Trends Over Time}
To analyze the trends of topics, we define the \textit{impact} \cite{anton} of a topic $z_{k}$ in month $m$ as,
\begin{equation}
impact(z_k, m) = \dfrac{1}{|D(m)|} \sum_{d_i\in D(m)} \theta(d_{i},z_{k})
\end{equation}
where $D(m)$ is the set of all posts in the month $m$. The \textit{impact} metric measures the relative proportion of posts related to that topic compared to the other topics in that particular month.

\section{Suggesting Tags}
Stack Overflow identifies each question posts with the tags which are given by the developers while posting the question. Unnecessary tags might be generated by users who does not have prior knowledge about the existing tags. To overcome this problem, when the user inputs a question, it is analyzed to find the latent topics and the tags are suggested based on those topics. 

\section{Quality Analysis of Questions in Stack Overflow}
The number of questions added each month to Stack Overflow has been steadily growing since the start of Stack Overflow and it has reached the maximum of more than 200,000 new questions per month. The quality of text content provided by the Stack Overflow website may vary and ranges from good quality questions/answers to low quality questions/answers. 
\par
In this work we have focused on different levels of quality of question posts in order to maintain the growth of Stack Overflow website. Since there are large amount of questions being posted each month, some of the questions may be answered and some remain unanswered. Thus by identifying the quality of questions in Stack Overflow in each topic discovered by LDA, using the score value of each question post, we can discard the low quality questions from Stack Overflow. We also discard the question posts whose score value is 0. We assume that 0-scored questions have not attracted enough interests from the community of developers. After removing 0-scored question, our dataset contains 94,879 questions which we subdivide into three categories:
\begin{enumerate}
\item Good Quality: Questions with accepted answers and score value greater than 7 fall into this category
\item Medium Quality: Questions with accepted answer and score value between 1 to 6 fall into this category
\item Low Quality: Questions with no accepted answer and score value less than 0 fall into this category.
\end{enumerate}

\section{Results and Discussion}
The 10 topics discovered by LDA and some of its top LDA words are shown in Table \ref{tab:t1}.

\begin{figure}[bpht!]
			\begin{center}
				\includegraphics[height=6.0cm]{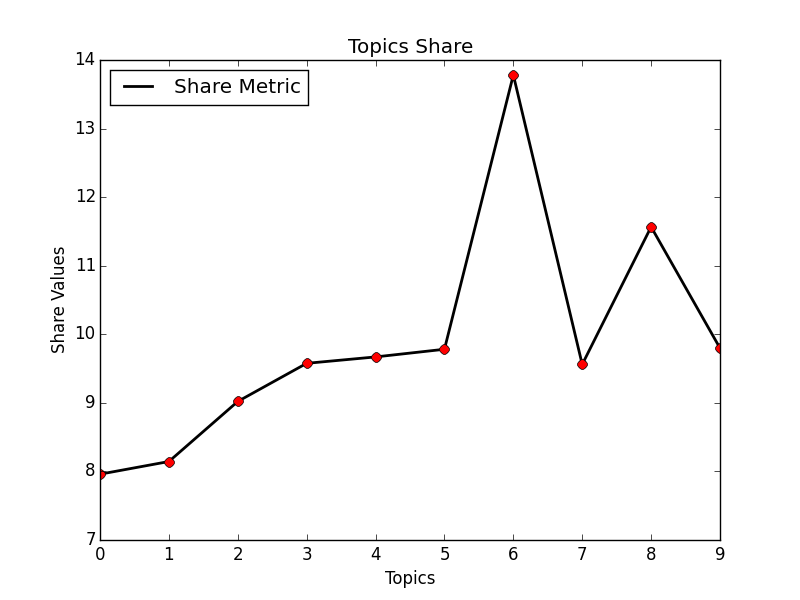}
				\caption{Topics Share Graph}
				\label{fig:f3}
			\end{center}
		\end{figure}

\begin{table}[bpht!]
		\caption{Topics and its Top Words}
		\label{tab:t1}
			\begin{center}
				\begin{tabular}{|c|c|}
						\hline\hline
						 Topic Names & Top LDA words \\ [0.5ex] 
						 \hline\hline
						 Topic 0 & memori thread alloc process pointer \\ 
						 \hline
						 Topic 1 & except error log messag fail \\
						 \hline
						 Topic 2 & think peopl question develop dai \\
						 \hline
						 Topic 3 & web net java http framework \\
						 \hline
						 Topic 4 & tabl sql queri insert updat \\ 
						 \hline
						 Topic 5 & python perl rubi php script \\
						 \hline
						 Topic 6 & string arrai declar argument list \\
						 \hline
						 Topic 7 & svn repositori control branch chang \\
						 \hline
						 Topic 8 & page button click css browser \\
						 \hline
						 Topic 9 & session secur password login site \\ [1ex]
						 \hline\hline
						\end{tabular}
				\end{center}	
			\end{table}
We have 10 topics discovered by TMT tool from Stack Overflow posts data. The topics are arranged in ascending or descending order according to their \textit{share} metric. Figure \ref{fig:f3} shows the \emph{share} graph for each topic.  
		
\begin{figure}[bpht!]
				  \begin{center}
					\includegraphics[height=6.0cm]{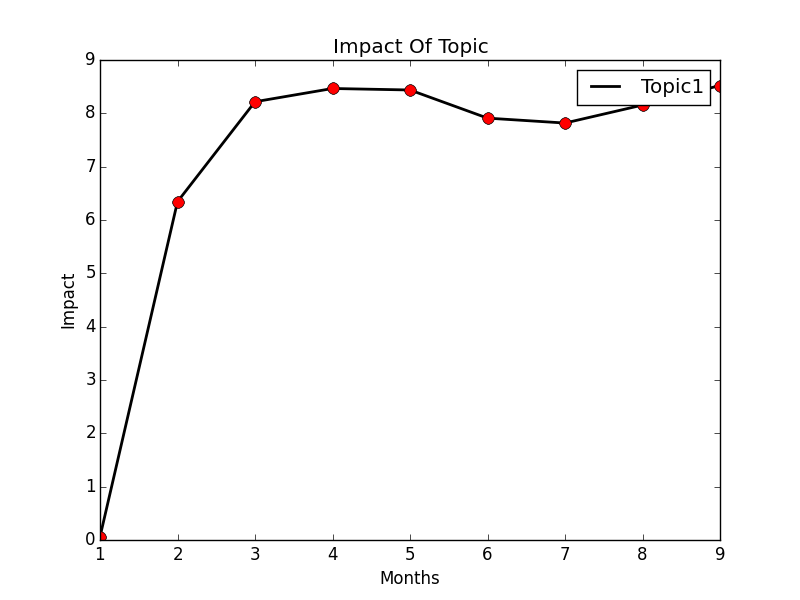}
					\caption{Impact Of Topic1}
					\label{fig:f4}
					\end{center}
	\end{figure}
From Fig.\ref{fig:f3} it is observed that the Topic6 is the most discussed topic among the 10 discovered topics in Stack Overflow. Higher the \emph{share} value, more the topic is being discussed in stack Overflow.
				
				\begin{figure}[bpht!]
				  \begin{center}
					\includegraphics[height=6.0cm]{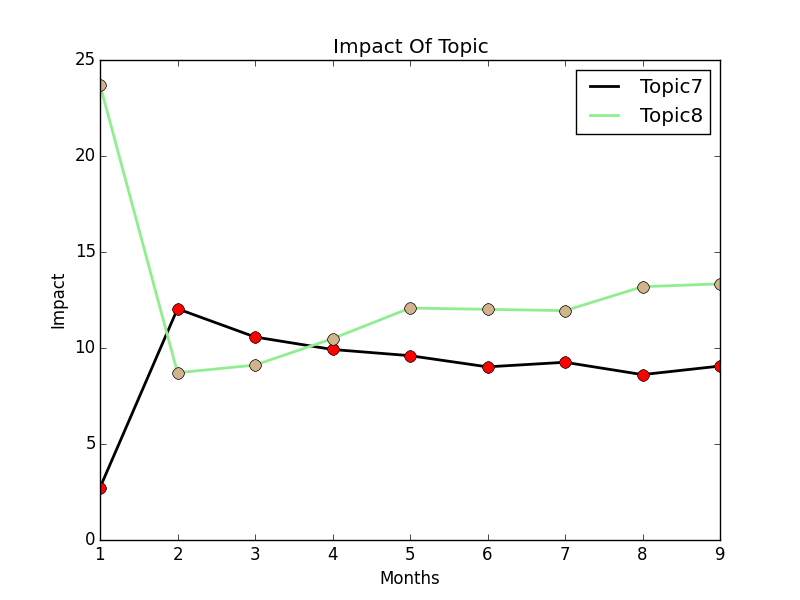}
					\caption{Impact Of Two Topics}
					\label{fig:f5}
					\end{center}
				\end{figure}

Figure \ref{fig:f4} and \ref{fig:f5} projects the trend line of each topic and comparison between them based on \textit{impact} values. These trend line indicates the rise or fall of interest in a particular topic on Stack Overflow. Table \ref{tab:t2} shows the overall result which contains topics discovered by LDA, \emph{share} value and the trend of each topic. 
		\begin{table}[bpht!]
		\caption{Topics Shares and Trends}
		\label{tab:t2}
		\begin{center}
					\begin{tabular}{|c|c|c|}
							\hline\hline
							 Topic Names & Share(\%) & Trend \\ [0.5ex] 
							 \hline\hline
							 Topic6 & 13.78 & $\Downarrow$ \\ 
							 \hline
							 Topic8 & 11.56 & $\Downarrow$ \\
							 \hline
							 Topic9 & 9.79 & $\Uparrow$ \\
							 \hline
							 Topic5 & 9.78 & $\Downarrow$ \\
							 \hline
							 Topic4 & 9.66 & $\Downarrow$ \\ 
							 \hline
							 Topic3 & 9.58 & $\Uparrow$ \\
							 \hline
							 Topic7 & 9.56 & $\Uparrow$ \\
							 \hline
							 Topic2 & 9.02 & $\Downarrow$ \\
							 \hline
							 Topic1 & 8.14 & $\Uparrow$ \\
							 \hline
							 Topic0 & 7.96 & $\Downarrow$ \\ [1ex]
							 \hline\hline
							\end{tabular}
					\end{center}
		\end{table}
 
Table \ref{tab:t3} tabulates the results of tag suggestion.
\begin{table}[bpht!]
		\caption{Tags Suggestion Result}
		\label{tab:t3}
		\begin{center}
					\begin{tabular}{|p{4.5cm}|p{1.5cm}|p{1.5cm}|}
							\hline\hline
							 Question & Topics & Tags \\ [0.5ex] 
							 \hline\hline
							 I have a Python program that works with dictionaries a lot. I have to make copies of dictionaries thousands of times. I need a copy of both the keys and the associated contents. The copy will be edited and must not be linked to the original (e.g. changes in the copy must not affect the original.) & Topic8, Topic7, Topic1, Topic5 & jquery,c\#, python, javascript, asp.net \\ 
							 \hline
							 I would like to know how to export in Arcgis a list of values calculated in python script into one of the following data formats: csv, txt, xls, dbase or other. I would also like to know how to create such file in case that it doesnt exist. & Topic6, Topic8, Topic4, Topic1 & mysql, sql-server, java, database \\
							 \hline
							 I have multiple tables in a database.I need to have a SQL statement in which i can get this values. How will i get this code in SQL using those join methods, im writing it in VB6 ADODC, is this the same syntax in a standard SQL. & Topic4 & database, sql, mysql, c\#, php \\
							 \hline\hline
							\end{tabular}
					\end{center}
		\end{table}

		\begin{figure}[bpht!]
				  \begin{center}
					\includegraphics[height=5.0cm]{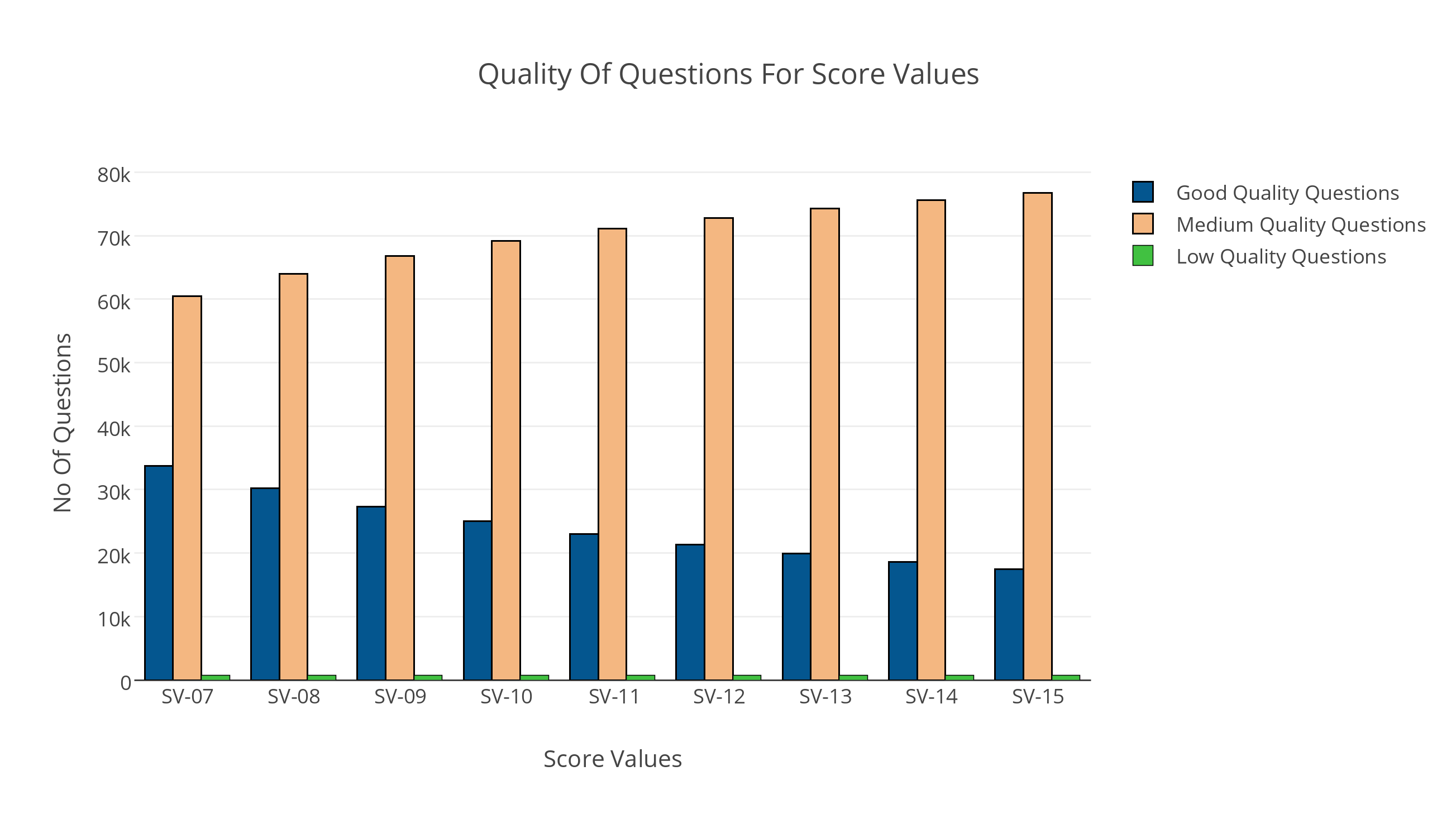}
					\caption{Quality of Questions for particular Score Values}
					\label{fig:f6}
					\end{center}
		\end{figure}
		
		\begin{figure}[bpht!]
						  \begin{center}
							\includegraphics[height=5.0cm]{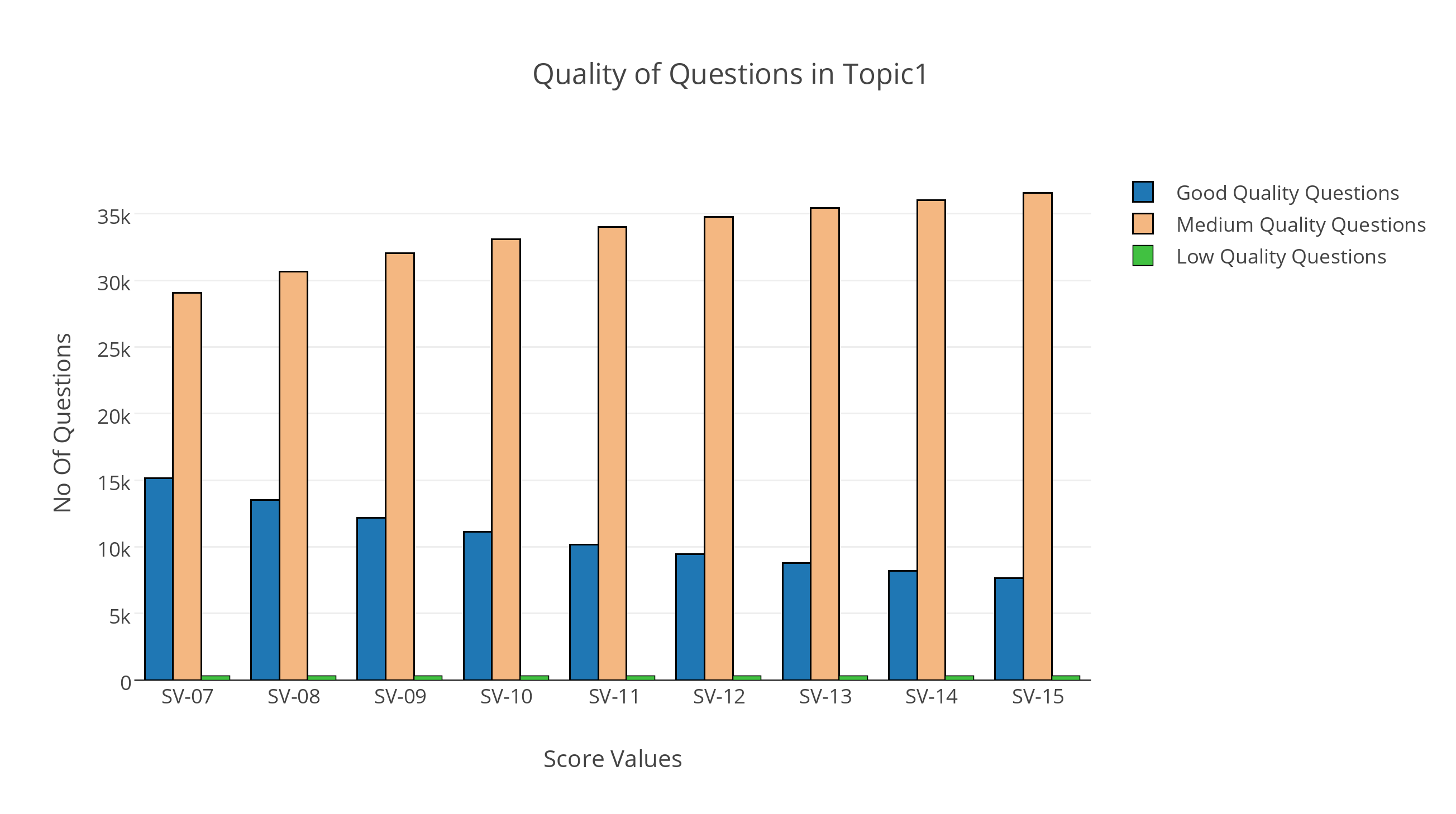}
							\caption{Quality of Questions in Topic1}
							\label{fig:f7}
							\end{center}
		\end{figure}
Figure \ref{fig:f6} depicts variation in number of quality question for a particular score value, and Fig.\ref{fig:f7} shows the variation in number of quality question for a particular topic, Topic1. 
\par
Every user face their own problems during programming. So they tend to obtain help from programming Q\&A websites like Stack Overflow. The posts belong to various topics and trend of these topics change over time, which needs to be analyzed before deciding if the posts are useful or not. Topic Modeling technique LDA can handle numerous latent topics that can be found in large set of data. The probability distribution obtained after applying LDA is reliable and gives expected results, which can be followed by analyzing the quality of questions in Stack Overflow and suggesting tags to the new questions. The proposed method help to eliminate the low quality question and at the same time suggest tags for a given question instead of simply adding more tags to Stack Overflow. This also helps to maintain the growth of Stack Overflow website.

\section{Conclusion}\label{conc}
In this paper, we proposed a method to identify the topics present in the Stack Overflow posts dataset. We also focused on the trend of those topics, how they change over time in order to understand the developer's discussion in Stack Overflow. Topic Modeling technique, LDA is used in our method to discover topics from the textual content of Stack Overflow posts dataset. This is followed by technique to suggest tags for the upcoming question posts to eliminate the generation of new and unnecessary tags. By doing this we can avoid overwhelming of tags within Stack Overflow. 
\par 
Since the generation of new questions are increasing each month, we analyzed the quality of questions to avoid the questions that are unanswered. We consider three different levels of quality which are categorized to Good Quality, Medium Quality and Low Quality questions. Good Quality include questions which have accepted answer and score value grater than 7. Medium Quality include questions having accepted answer and score value between 1 to 6. Low Quality include questions having score value less than 0. As we increase/decrease the score value, we identify the variation in quality of questions in Stack Overflow. 
\par 
Our proposed method helps to discard the low quality questions and unwanted tags and also recommend tags to new questions. This further helps in maintaining the growth of Stack Overflow website by removing unnecessary data from the website.   
 
\bibliographystyle{IEEEtran}
\bibliography{ref}

\begin{thebibliography}{10}
\providecommand{\url}[1]{#1}
\csname url@samestyle\endcsname
\providecommand{\newblock}{\relax}
\providecommand{\bibinfo}[2]{#2}
\providecommand{\BIBentrySTDinterwordspacing}{\spaceskip=0pt\relax}
\providecommand{\BIBentryALTinterwordstretchfactor}{4}
\providecommand{\BIBentryALTinterwordspacing}{\spaceskip=\fontdimen2\font plus
\BIBentryALTinterwordstretchfactor\fontdimen3\font minus
  \fontdimen4\font\relax}
\providecommand{\BIBforeignlanguage}[2]{{%
\expandafter\ifx\csname l@#1\endcsname\relax
\typeout{** WARNING: IEEEtran.bst: No hyphenation pattern has been}%
\typeout{** loaded for the language `#1'. Using the pattern for}%
\typeout{** the default language instead.}%
\else
\language=\csname l@#1\endcsname
\fi
#2}}
\providecommand{\BIBdecl}{\relax}
\BIBdecl

\bibitem{Mamykina}
\BIBentryALTinterwordspacing
L.~Mamykina, B.~Manoim, M.~Mittal, G.~Hripcsak, and B.~Hartmann, ``Design
  lessons from the fastest {Q}\&{A} site in the west,'' in \emph{Proceedings of
  the SIGCHI Conference on Human Factors in Computing Systems}, ser. CHI
  '11.\hskip 1em plus 0.5em minus 0.4em\relax New York, NY, USA: ACM, 2011, pp.
  2857--2866. [Online]. Available:
  \url{http://doi.acm.org/10.1145/1978942.1979366}
\BIBentrySTDinterwordspacing

\bibitem{Ponzanelli}
\BIBentryALTinterwordspacing
L.~Ponzanelli, A.~Mocci, A.~Bacchelli, and M.~Lanza, ``Understanding and
  classifying the quality of technical forum questions,'' in \emph{Proceedings
  of the 2014 14th International Conference on Quality Software}, ser. QSIC
  '14.\hskip 1em plus 0.5em minus 0.4em\relax Washington, DC, USA: IEEE
  Computer Society, 2014, pp. 343--352. [Online]. Available:
  \url{http://dx.doi.org/10.1109/QSIC.2014.27}
\BIBentrySTDinterwordspacing

\bibitem{Blei2}
\BIBentryALTinterwordspacing
D.~M. Blei, A.~Y. Ng, and M.~I. Jordan, ``Latent dirichlet allocation,''
  \emph{J. Mach. Learn. Res.}, vol.~3, pp. 993--1022, Mar 2003. [Online].
  Available: \url{http://dl.acm.org/citation.cfm?id=944919.944937}
\BIBentrySTDinterwordspacing

\bibitem{GGPG}
Z.~Gyongyi, G.~Koutrika, J.~Pedersen, and H.~Garcia-Molina, ``Questioning
  yahoo! answers,'' [Available Online]
  \url{http://infolab.stanford.edu/~zoltan/publications/gyongyi2008questioning.pdf},
  2008.

\bibitem{Adamic}
\BIBentryALTinterwordspacing
L.~A. Adamic, J.~Zhang, E.~Bakshy, and M.~S. Ackerman, ``Knowledge sharing and
  yahoo answers: Everyone knows something,'' in \emph{Proceedings of the 17th
  International Conference on World Wide Web}, ser. WWW '08.\hskip 1em plus
  0.5em minus 0.4em\relax New York, NY, USA: ACM, 2008, pp. 665--674. [Online].
  Available: \url{http://doi.acm.org/10.1145/1367497.1367587}
\BIBentrySTDinterwordspacing

\bibitem{Treude}
\BIBentryALTinterwordspacing
C.~Treude, O.~Barzilay, and M.-A. Storey, ``How do programmers ask and answer
  questions on the web?'' in \emph{Proceedings of the 33rd International
  Conference on Software Engineering}, ser. ICSE '11.\hskip 1em plus 0.5em
  minus 0.4em\relax New York, NY, USA: ACM, 2011, pp. 804--807. [Online].
  Available: \url{http://doi.acm.org/10.1145/1985793.1985907}
\BIBentrySTDinterwordspacing

\bibitem{Bajracharya}
\BIBentryALTinterwordspacing
S.~K. Bajracharya and C.~V. Lopes, ``Analyzing and mining a code search engine
  usage log,'' \emph{Empirical Softw. Engg.}, vol.~17, no. 4-5, pp. 424--466,
  Aug 2012. [Online]. Available:
  \url{http://dx.doi.org/10.1007/s10664-010-9144-6}
\BIBentrySTDinterwordspacing

\bibitem{Pagano}
\BIBentryALTinterwordspacing
D.~Pagano and W.~Maalej, ``How do open source communities blog?''
  \emph{Empirical Softw. Engg.}, vol.~18, no.~6, pp. 1090--1124, Dec 2013.
  [Online]. Available: \url{http://dx.doi.org/10.1007/s10664-012-9211-2}
\BIBentrySTDinterwordspacing

\bibitem{stemming}
Wikipedia, ``Stemming --- wikipedia{,} the free encyclopedia,''
  \url{http://en.wikipedia.org/w/index.php?title=Stemming&oldid=653890337},
  2015, [Online; accessed 2-April-2015].

\bibitem{Blei1}
\BIBentryALTinterwordspacing
D.~M. Blei, ``Probabilistic topic models,'' \emph{Commun. ACM}, vol.~55, no.~4,
  pp. 77--84, April 2012. [Online]. Available:
  \url{http://doi.acm.org/10.1145/2133806.2133826}
\BIBentrySTDinterwordspacing

\bibitem{Lecture1}
------, ``Topic models,'' University Lecture, November 2009.

\bibitem{Lecture2}
M.~Sklar, ``Digging into the dirichlet distribution,''
  \url{https://www.hakkalabs.co/articles/the-dirichlet-distribution}, December
  2013, lecture given at New York Machine Learning meetup at Pivotal Labs.

\bibitem{plate}
Wikipedia, ``Latent dirichlet allocation --- wikipedia{,} the free
  encyclopedia,''
  \url{http://en.wikipedia.org/w/index.php?title=Latent_Dirichlet_allocation&oldid=650781080},
  2015, [Online; accessed 2-April-2015].

\bibitem{Lecture}
J.~Speh, A.~Muhic, and J.~Rupnik, ``Parameter estimation for the latent
  dirichlet allocation,'' [Available Online]
  \url{http://ailab.ijs.si/dunja/SiKDD2013/Papers/Speh-ldaAlgorithms.pdf}, Oct
  2013.

\bibitem{tmt}
T.~S. N. L.~P. Group, ``Standford topic modeling toolbox,'' [Available Online]
  \url{nlp.stanford.edu/software/tmt/tmt-0.4/}, 2010.

\bibitem{anton}
\BIBentryALTinterwordspacing
A.~E.~H. Anton~Barua, Stephen W.~Thomas, ``What are developers talking about?
  an analysis of topics and trends in stack overflow,'' \emph{Empirical
  Software Engineering}, vol.~19, no.~3, pp. 619--654, June 2014. [Online].
  Available:
  \url{http://link.springer.com/article/10.1007\%2Fs10664-012-9231-y}
\BIBentrySTDinterwordspacing

\end{thebibliography}
\end{document}